\documentclass[reprint,amssymb, amsmath, aip,cha,twocolumn]{revtex4-1}
\usepackage{graphicx}
\usepackage[caption = false]{subfig}
\usepackage{color}
\usepackage{epsfig}
\usepackage{ifpdf}
\usepackage{url}%
\usepackage{bm}
\usepackage{soul}
\usepackage[colorlinks=true,linkcolor=blue]{hyperref}%
\usepackage{cleveref}
\usepackage{soul}
\expandafter\ifx\csname package@font\endcsname\relax\else
 \expandafter\expandafter
 \expandafter\usepackage
 \expandafter\expandafter
 \expandafter{\csname package@font\endcsname}%
\fi
\begin{document}
\title{Thermodynamics and Self-organization of Strongly Coupled Coulomb Clusters: An Experimental Study}
\author{MG Hariprasad}%
\email{hari.prasad@ipr.res.in}
\author{P. Bandyopadhyay}
\author{ Garima Arora}
\author{A. Sen}
\affiliation{Institute For Plasma Research, HBNI, Bhat, Gandhinagar, Gujarat, India, 382428}%
\date{\today}
\begin{abstract}
In this experimental work, the thermodynamics and self-organization of classical two-dimensional Coulomb clusters are studied as a function of the cluster size. The experiments are carried out in a DC glow discharge Argon plasma in the Dusty Plasma Experimental (DPEx) device for clusters with different number of particles. Hexagonal symmetry around each individual particle is quantified using the local orientational order parameter ($|{\psi_6}|$) for all the configurations. The screened Coulomb coupling parameter, which plays a key role in determining the thermodynamic nature of a Coulomb cluster, is estimated using Langevin dynamics and found to be sensitive to the number of particles present in the cluster. In addition, the process of self-organization and the dynamics of individual particles of the cluster as it changes from a metastable state to the ground state are examined through the estimation of dynamic entropy. Our findings suggest an intimate link between the configurational ordering and the thermodynamics of a strongly coupled Coulomb cluster system - an insight that might be of practical value  in analysing and controlling  the micro dynamics of a wider class of finite systems.

\end{abstract}
\maketitle
\section{ Introduction}
Cluster science has been a fascinating and active field of research over the last few decades due to its diverse applications in nanotechnologies and colloidal science \cite{clusterscience1,clusterscience2,clusterscience3,clusterscience4,clusterscience6,clusterscience5}.  On account of its high surface to volume ratio,  a finite cluster shows anomalous physical and chemical behaviors in comparison to  bulk materials.  For example,  at the nano-scale, gold clusters are no longer fully noble and non-reactive and have applications in gas sensing, pollution reduction and catalysis \cite{goldcluster}.  Likewise,  the energy and entropy of  sodium liquid and solid clusters \cite{sodiumcluster}, are found to fluctuate with the number of particles (135$\le N \le$360) and the shell structures. In chemical science, clusters with four to five atoms of metals are normally used for catalytic activities and the activity is found to be highly sensitive to the number of constituent atoms in the cluster \cite{chemistry}.  Clusters may be electrically charged or neutral and can have various interactive potentials depending upon the nature of the cluster and the constituent components\cite{sodiumcluster}. 
As examples, NaCl clusters display strong attraction, Na and Cu clusters show ionic bonding , He and Ar clusters are subject to Van der Waals attractive forces  and Si clusters experience forces due to covalent chemical bonding \cite{clusterscience6}. Moreover, the design of two and three-dimensional clusters and their self-assembly play a major role in the fabrication of new functional materials and electronic devices.\cite{functional2,functional3,functional4}. \par
 Colloidal systems \cite{colloid1,colloid2,colloid3} and complex plasmas  \cite{pkshukla,goreerfanddc,plasmacrystalmelting,morfill2009complex} serve as convenient platforms  for cluster studies and have been widely employed  as model systems to investigate the dynamics of classical charged clusters at the micro-scale and at individual particle levels. A complex plasma is a multi-component system with neutrals, electrons, ions and highly charged massive dust particles. It is possible to manipulate a complex plasma 
 in a laboratory device to create  one\cite{zigzag}, two \cite{modespectratwod} or three-dimensional \cite{phasetransition2} finite classical Coulomb clusters. The dynamics of such clusters can be investigated in detail at the individual  particle level through optical diagnostics\cite{Vladimirov}. Melzer \textit{et.al}. \cite{review} have provided an  excellent  review of past research on such Coulomb clusters highlighting the role of different  collective modes in clusters and the effect of ion wakes on cluster dynamics. The continuous and discontinuous phase transitions of a two-dimensional cluster with only six constituent particles, have been investigated by Sheridan \cite{phasetransition1}.   
 The melting of Coulomb balls with N=32 and N=35 were studied by Schella {\it et al.} \cite{phasetransition2} using two different methods. In the first method the cluster is melted by increasing the plasma power which drives a non-equilibrium melting due to ion wakes while  in the second method the cluster is melted using laser power that is akin to an equilibrium scenario. The mode spectra of finite Coulomb clusters have also been investigated to explore cluster phase transitions and for estimation of dusty plasma parameters  \cite{modespectra,zigzag,modespectratwod}.  
 
  The transition of a three-particle cluster to chaotic dynamics was studied by Sheridan \textit{et.al} \cite{chaos}   by employing a periodic driving force with two different driving frequencies and measuring the mode spectra. Klindworth {\it et al} \cite{laserexcited} experimentally  investigated  laser-excited inter-shell rotation induced structural transitions of finite Coulomb clusters and their dynamical stability by  applying a well defined torque on the cluster using the radiation pressure of two opposing laser beams . The dynamic and thermodynamic properties during the non-equilibrium two-step melting of a finite Coulomb cluster was  addressed in great detail by Ivanov and Melzer \cite{noneqmelting}. It was found that as the background neutral pressure was decreased a single particle beneath a complex plasma cluster melted the cluster through ion wake mediated interactions. 

 By adopting the concept of mean first-passage time dynamic entropy \cite{dynamicentropyref} of each dust particle, the phase transition in the Coulomb cluster of strongly interacting Brownian particles has been investigated very recently\cite{dynamicentropy} in crystalline, liquid, and transient states. Dynamic entropy of each particle is estimated for different kinetic temperatures. As an extension, the same group has studied the behavior of an active matter Coulomb cluster \cite{activematter}. \par

Most of these past studies on clusters have been devoted to clusters of a fixed size  and have focused on the energetics, thermodynamics, growth kinetics and the dependency of the cluster properties on the interacting potential between the constituent components \cite{clusterscience6}. There seems to have been no systematic experimental study on the  thermodynamic properties of a cluster as a function of the number of particles in the cluster. Likewise the dynamics of the individual particles during the process of self-organization of the clusters while the system transits from an excited state to the ground state, has not received much attention.   In this paper we report our experimental investigation of the variation of the thermodynamical parameters of Coulomb clusters as the number of constituent  particles is changed. We use  the Langevin dynamics method for investigating clusters with different number of particles. The structural hexagonal symmetry of the individual cluster components is examined by estimating the local orientational order parameter ($|{\psi_6}|$). 
Furthermore, the dynamic entropy and the dynamics of each individual component is estimated during the self-organization process of the cluster during its transition from a metastable state to the ground state. The particles which are stable in their respective equilibrium positions and which are re-arranging are identified from the dynamic entropy analysis.  \par
The paper is organized as follows: Section~\ref{sec:setup} explains the experimental set-up and describes the experimental procedures. Section~\ref{sec:results} contains the experimental observations and  relevant discussions on them. A brief summary and some concluding remarks  are provided in section~\ref{sec:conclusion}.

\section{Experimental Set-up}
\label{sec:setup}
\begin{figure}[ht]
\includegraphics[scale=0.95]{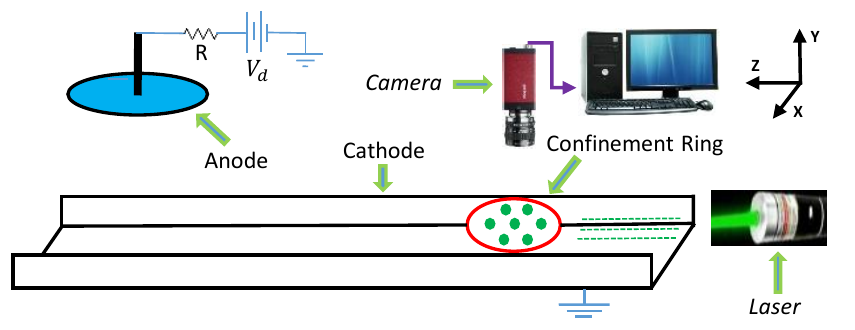}
\caption{\label{fig:fig1} Schematic diagram of Dusty Plasma Experimental (DPEx) device. $V_d$ indicates the voltage of the current source and $R$ represents the resistance of the current limiting resistor connected in series with the current source.}
\end{figure}
The present set of experiments is carried out in the Dusty Plasma Experimental (DPEx) device whose schematic diagram is presented in Fig.~\ref{fig:fig1} and whose detailed description can be found in reference \cite{dpex}. After achieving a base pressure of $10^{-3}$ mbar using a rotary pump, Argon gas is injected into the chamber to set the working pressure to 0.1 mbar.  A glow discharge plasma is then produced by applying a DC voltage of $V_d= 300$~V across the non-symmetric electrodes as depicted in Fig.~\ref{fig:fig1} and the discharge current ($I_d=2$~mA) is obtained by measuring the voltage drop across a current limiting resistor of resistance $R=2$~k$\Omega$, which is connected in series with the power supply to limit the flow of current in the circuit. After producing the plasma, mono-dispersive Melamine Formaldehyde dust particles of diameter 10.66 $\pm$ 0.01 $\mu m$ are introduced into the chamber from the top of the cathode. While falling down, these massive dust particles attain a high negative charge because of the higher mobility of electrons compared to the ions.  The charged particles levitate near the cathode sheath region due to a balance between  the downward gravitational force and the upward electrostatic force due to the sheath electric field. As a result they form an ordered structure in the x-z plane. The first observation of such a Coulomb cluster in the DPEx device was recently reported by Hariprasad \textit{et al.} \cite{dpexcrystal}. The horizontal confinement of these mutually repelling dust particles is attained by the sheath electric field produced by a circular SS ring of diameter 5 cm as shown in Fig.~\ref{fig:fig1}.  A green semiconductor diode laser is used to illuminate the dust particles, and a CCD camera is used to capture the still images of dust particles. To investigate the dynamics of this two-dimensional complex plasma structure, IDL and Matlab based softwares are used to analyze the images which are stored in a high-speed computer.
\section{ Results and Discussions}\label{sec:results}
\subsection{Addition of particles in Coulomb clusters}
\begin{figure}[ht]
\includegraphics[scale=0.27]{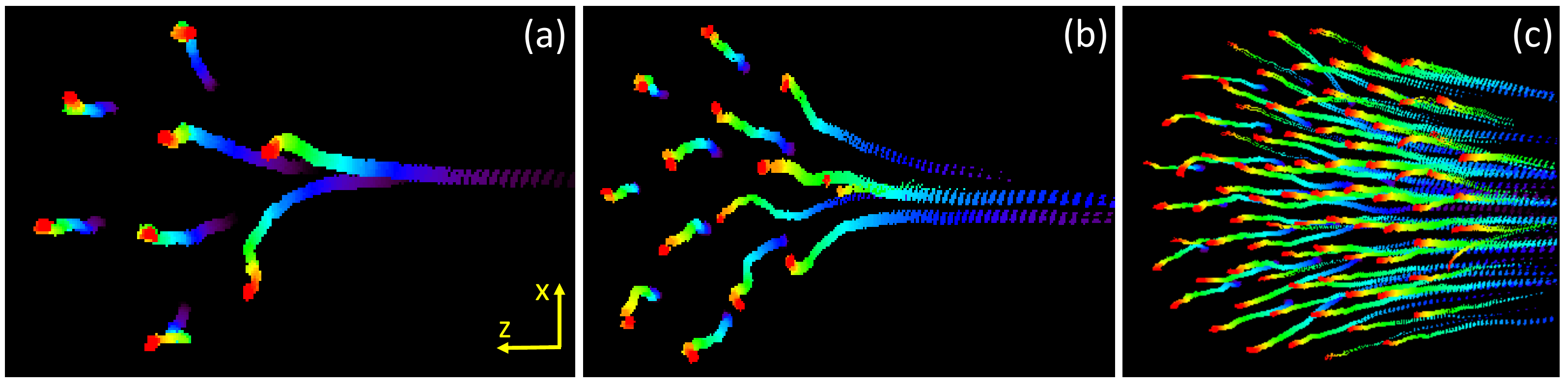}
\caption{\label{fig:fig2}Trajectories of dust particles while they enter the dust cluster. Colors represent the progress of  time with blue representing the initial time and red the final time. (a) Addition of 2 particles to a cluster with N=6 (b) Addition of 5 particles to a cluster with N=9 (c) Addition of $\sim$ 40 particles to
a cluster with N$\sim$30}
\end{figure}

As explained in Sec. \ref{sec:setup}, a small dust cluster is initially produced with a number of particles for a given set of discharge parameters $p=$0.1 mbar and $V_d=$ 300 V. Thereafter, a few more particles are added in steps to the cluster which is then seen to rearrange itself to a new ground state (stable equilibrium). The ground state configurations of dusty plasma clusters are determined both by the number of particles and the ambient plasma parameters \cite{totsuji,Kong}. In all the sets of our experiments, the plasma parameters are kept constant while varying the number of particles as N = 9, 14, 18, 19 and 28 and the resultant ground states examined. 
 Fig.~\ref{fig:fig2} shows the overlapping positions of the constituent particles in a cluster along with newly added particles over a number of image frames. The color scheme indicates the passage of time with blue representing the initial time and red the final time. 
Since the negatively charged dust particles are in a plasma environment the mutual interaction between the particles shown in  Fig.~\ref{fig:fig2} are through a shielded Coulomb (Yukawa) potential \cite{Uweprl,totsuji,review}.
 In Fig.~\ref{fig:fig2}(a), two dust particles have been added to a cluster of N=6. The added particles are seen to follow straight-line trajectories till they come close and get deflected thereafter as they get close to the cluster.  The particles within the cluster also get perturbed by the oncoming new particles and together they undergo a process of self-organization to form a new ground state cluster configuration. 
A similar scenario plays out in Fig.~\ref{fig:fig2}(b) and Fig.~\ref{fig:fig2}(c)  during the addition of particles to bigger clusters. A detailed analysis of the individual particle dynamics during the the transition to a new ground state is  presented in the upcoming section (Sec. \ref{sec:Entropy}).  

 \subsection{Local orientational order parameter ($|{\psi_6}|$)}
\begin{figure*}[ht]
\includegraphics[scale=0.45]{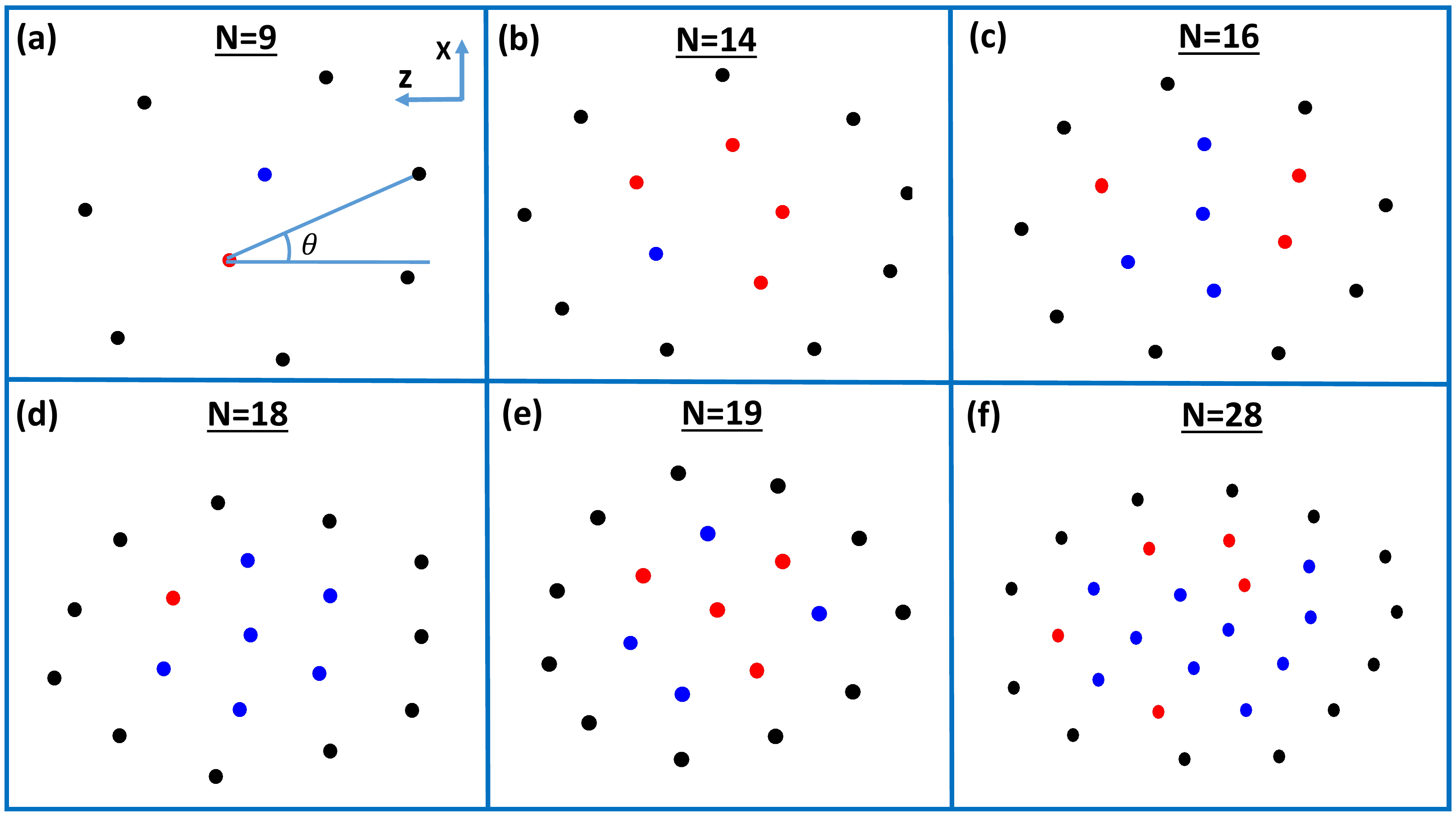}
\caption{\label{fig:fig3}Representation of $|\psi_6|$ for clusters with different numbers of particles: a) $N=9$, b) $N=14$, c) $N=16$, d) $N=18$, e) $N=19$ and f) $N=28$. The blue dots depict the particles with $|\psi_6| \geq 0.6$ whereas the red dots correspond to the particles with $|\psi_6| < 0.6$. The black dots represent the particles whose $|\psi_6|$ are not evaluated. }
\end{figure*}

In two-dimensional structures, the hexagonal ordering in the system can be quantified by measuring the local orientational order parameter of each particle \cite{Bonitzbook}. The local orientational order parameter ($|{\psi_6}|$) \cite{Bonitzbook} is defined as $|\Psi_6({\mathbf{r_j}})| = \left| \frac{1}{c_j} \sum_{1}^{nn(j)} e^{i6\theta_j}\right|$, where, $c_j$ is the coordination number of the \textit{j}th particle and $nn(j)$ are the nearest neighbours to this \textit{j}th particle. $\theta_j$ is the angle made by the Z-axis and the line which connects the centres of the particles under consideration ($j$th particle) and the neighbouring particles as depicted in Fig.~\ref{fig:fig3}(a). Basically, the local orientational order parameter ($|{\psi_6}|$) estimates the extent of hexagonal symmetry around a reference particle. For a  two-dimensional crystal with perfect hexagonal symmetry, $|{\psi_6}|$ becomes equal to one and its value diminishes with the deformation of the crystal \cite{Bonitzbook}. \par
In our experiments, $|{\psi_6}|$ is estimated for every individual particle for a different cluster with a various number of particles and is presented in Fig.~\ref{fig:fig3}. It is to be noted that while estimating $|{\psi_6}|$, the coordination number ($c_j$) and the nearest neighbors ($nn(j)$) are calculated from the Delaunay triangulation and Voronoi diagram. As the Voronoi diagram cannot be constructed for those particles that reside at the outermost shell, therefore  $|{\psi_6}|$ is estimated only for the particles which reside inside the cluster. The blue dots in Fig.~\ref{fig:fig3} represent the particles with $|{\psi_6}| \geq 0.6$ whereas the red dots correspond to the particles with $|{\psi_6}| < 0.6$. The black dots represent the particles whose $|{\psi_6}|$ are not evaluated. $|{\psi_6}|$ is found to be highly sensitive to the number of particles in the cluster and even the introduction of one particle completely distorts the symmetry of the system. Fig.~\ref{fig:fig3}(a) presents $|{\psi_6}|$ for a cluster made of nine particles. It indicates that there exists only one particle in the cluster having $|{\psi_6}| \ge 0.6$ and the number of particles is not sufficient to form more than one hexagonal cell in the cluster. For $N=14$ (see Fig.~\ref{fig:fig3}(b)) also, only one blue dot is visible in the cluster. The hexagonal symmetry becomes better with the addition of two more particles to a dust cluster of $N=14$ as depicted in Fig.~\ref{fig:fig3}(c). Furthermore, the most stable state is achieved for $N=18$ and the cluster evolves to a state with nicely ordered cells and is shown in Fig.~\ref{fig:fig3}(d). There are 6 particles out of 7 in the inner shell with higher hexagonal symmetry ($|{\psi_6}| \ge 0.6$). The scenario changes dramatically with the addition of just one particle in a cluster having 18 particles. In this case, the hexagonal symmetry of the cluster is destroyed, and only four out of eight particles survive with hexagonal symmetry (see Fig.~\ref{fig:fig3}(e)). Instead of forming ordered cells, the newly added particle displaces another particle in between the two shells. The local orientational order parameter for a cluster with 28 particles is shown in Fig.~\ref{fig:fig3}(f). In this case, ten particles in the inner cells are found to have $|{\psi_6}| \ge 0.6$ and five particles do not possess hexagonal symmetry. A detailed analysis of the thermodynamical properties of a cluster with different number of particles will be discussed in Sec.~IIIC.

\begin{figure*}[ht]
\includegraphics[scale=0.80]{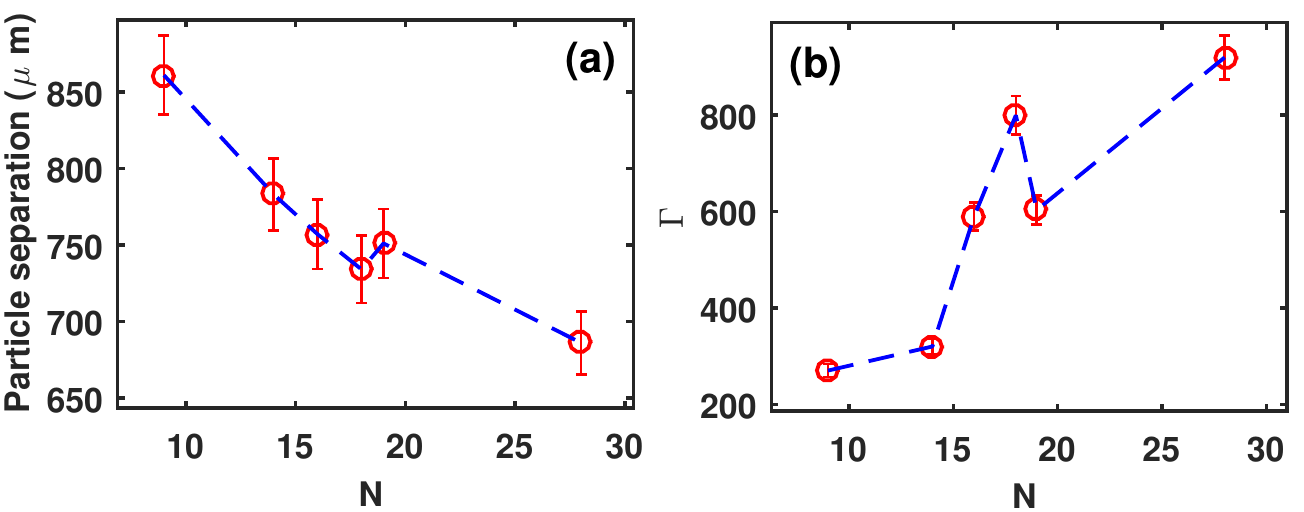}
\caption{\label{fig:fig4} Variation of (a) Inter-particle distance   (b) Screened Coulomb coupling parameter with different number of particles.}
\end{figure*}

\subsection{Thermodynamics of dust clusters}\label{sec:thermo}
The thermodynamics of a system is always of interest since it determines the phase state of the system. However, determining the thermodynamical parameters of a few particle system  is very challenging. In our experiment, we investigate the  thermodynamics of dust clusters  by estimating the Coulomb coupling parameter \cite{gammathermodynamic} through Langevin dynamics \cite{langevindynamics,dpexcrystal}. As has been discussed in ref \cite{gammathermodynamic} the coupling parameter provides a good measure of the thermodynamic state of a system . 
Using Langevin dynamics, the Coulomb coupling parameter can be estimated directly using the position information over time of all the particles of a system without a knowledge of plasma and dusty plasma parameters such as dust charge, dust temperature, electron temperature, plasma density and interparticle distance. According to the method of Langevin dynamics, the probability to find a particle at position {\lq r\rq}  with a velocity of {\lq v\rq} is given as  \cite{langevindynamics,dpexcrystal}	 
\begin{eqnarray}
P(r,v)\propto exp\left[-\frac{m{(v-<v>)}^2}{2T}-\frac{m{\Omega_E}^2r^2}{2T}\right],
\end{eqnarray}
with $T$ being the temperature of dust particles, $\Omega_E$ is the Einstein frequency and $m$ represents the mass of the particle.  The standard deviation of the displacement distribution yields the coupling parameter. The displacement distribution primarily provides the Einstein frequency which can be used to calculate the screened Coulomb coupling parameter once the inter-particle distance is known. The standard deviation of the displacement distribution is given by ${\sigma_r}= \sqrt{\frac{\Delta^2}{\Gamma_{eff}}}$. In the present set of experiments, the Coulomb coupling parameter is estimated from the knowledge of the inter-particle distance ($\Delta$) and the standard deviation of displacement distribution ($\sigma_r$) as $\Gamma_{eff}=\left[\frac{\Delta}{\sigma_r}\right]^2$. The average inter-particle distance ($\Delta$) of the cluster is estimated by taking the average separation between nearby particles from the information of position coordinates of individual particles.  In Fig.~\ref{fig:fig4}(a), the variation of inter-particle distance is plotted against the number of particles in the cluster. The inter-particle distance is found to decrease with an increase in the number of particles since the particles are confined in an electrostatic potential trap created by the sheath electric field and the extent of the confining region is limited. 

 Even though the inter-particle distance is decreasing with the number of particles, there is an anomalous behaviour when the number of particles becomes 19. In that case, the inter-particle distance increases as depicted in Fig.~\ref{fig:fig4}(a). Inter-particle distance reduces further when more particles are added to the cluster. This anomalous behaviour in inter-particle distance for the cluster with N=19 may be because of the lower hexagonal symmetry, since the hexagonal structures in the cluster closely pack the system with minimal energy. Just the introduction of one particle reduces the number of particles with hexagonal symmetry from six to four for N=18 and N=19 and this gets  reflected in the inter-particle distance also.  \par
  Moreover, the number of particles drastically affects the Coulomb coupling parameter ($\Gamma$) as shown in Fig.~\ref{fig:fig4}(b). The Coulomb coupling parameter is observed to follow the reverse trend of inter-particle distance. As can be seen from Fig. ~\ref{fig:fig4}(b), $\Gamma$  starts to increase with the number of particles, reaches the value of $\Gamma$=800 at N=18 then there is a dip in the value at N=19 which is the same as the anomalous behavior in the plot of inter-particle distance. This trend of $\Gamma$ can be explained from its dependence on inter-particle distance (\lq d') as $\Gamma \propto 1/(d\times\exp(d))$ \cite{gamma} which is qualitatively  based on the breaking of hexagonal symmetry with addition of one particle.  \par
In short, $\Gamma$, the parameter which determines the thermodynamic nature of the two-dimensional cluster is found to be dependent on its spatial configuration. This observation may have important practical implications {\it e.g.} in the context of functional materials such as colloidal micro and nano-particle clusters \cite{functional2,functional4}. The basic building blocks of these special materials include micro and nano-sized colloidal particles which are arranged in specific spatial orientation to achieve special properties in electronic and optoelectronic devices \cite{functional1,functional2}. A large research effort is directed to the fabrication of functional materials using these organized particles \cite{functional2,functional3,functional4}. A complex plasma system consists of charged micron-sized dust particles, which are arranged to form ordered or disordered structures and their dynamics has a close resemblance to the colloidal-particle assemblies. Hence, our observation regarding the connection between cluster thermodynamics and spatial ordering of complex plasma clusters may prove useful in the design and fabrication of such functional materials of colloidal-particle assemblies, where the ordering of building blocks and associated dynamics are critical parameters determining their functional behavior.

\subsection{Cluster Self-organization and Dynamic Entropy}\label{sec:Entropy}
\begin{figure}[ht]
\includegraphics[scale=0.81]{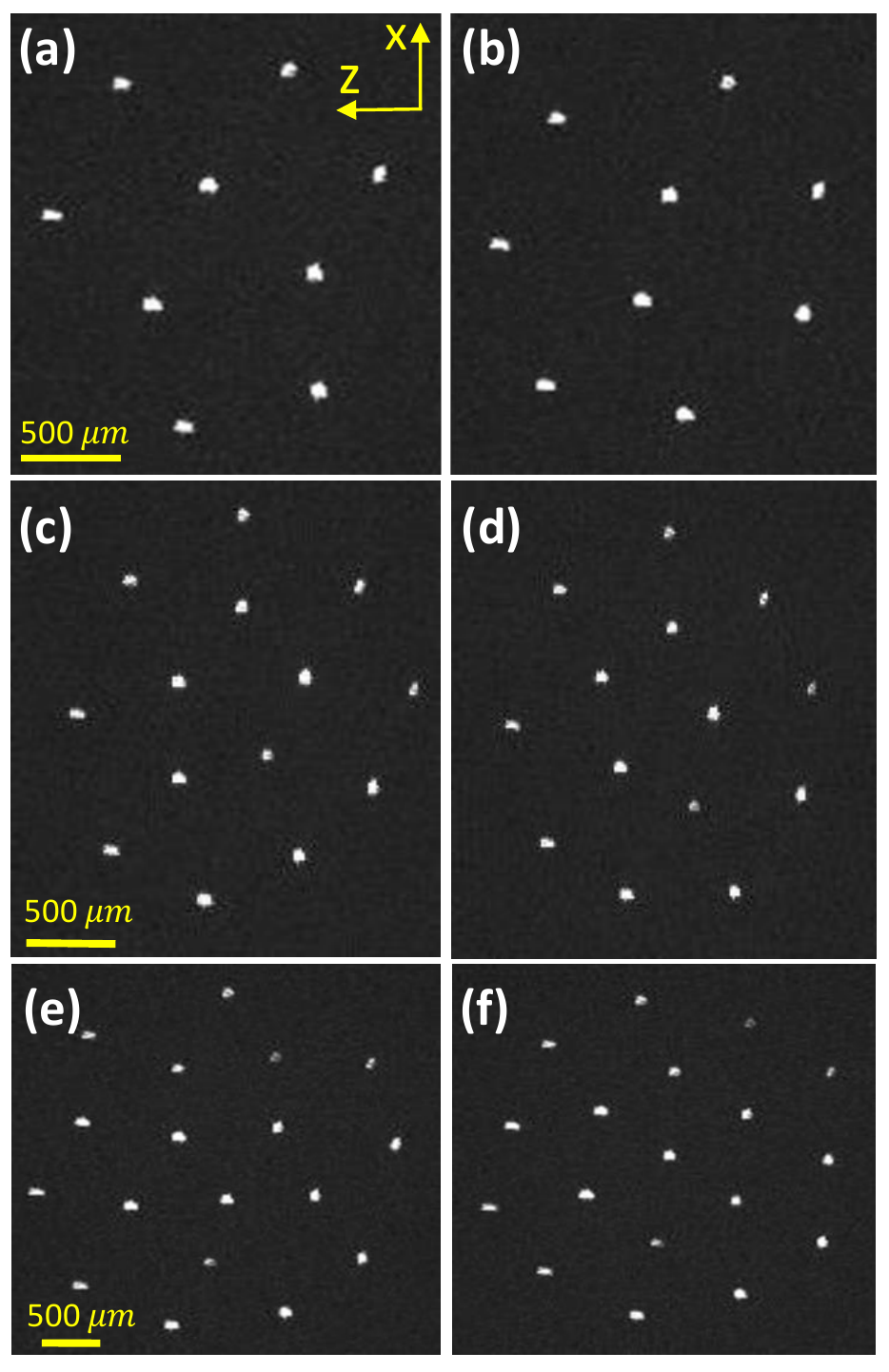}
\caption{\label{fig:fig5}  (a), (c), (e) Meta-stable states and (b), (d), (f) Ground states for clusters with different number of particles N=9, 14, 18, respectively.  }
\end{figure}

Dust clusters exhibit different configurations or ground states depending upon the plasma parameters \cite{totsuji,Kong}. According to past studies, the dust cluster either gets temporarily excited to a meta-stable state from the ground state \cite{metastable1,finitegamma} due to any kind of perturbation in the system or it simply passes through a meta-stable state during the formation of the equilibrium dust cluster.  In our experiments, such kind of excited metastable states are observed during the cluster formation. When dust particles are added to the cluster, it is found to be in an unstable configuration for some time and then it self-organizes to a new configuration which is stable over time. As the discharge parameters are kept constant in our experiments, hence one should expect there will be only one ground state for a given number of particles.  Fig.~\ref{fig:fig5} shows various meta-stable states (Fig.~\ref{fig:fig5} (a), (c) and (e)) and ground states  (Fig.~\ref{fig:fig5} (b), (d) and (f)) of dust clusters for different number of particles N=9, 14 and 18 respectively. It is worth mentioning that the ground states are dominated by hexagonal cells and they occupy a smaller area compared to the respective meta-stable states.
\begin{figure}[ht]
\includegraphics[scale=0.28]{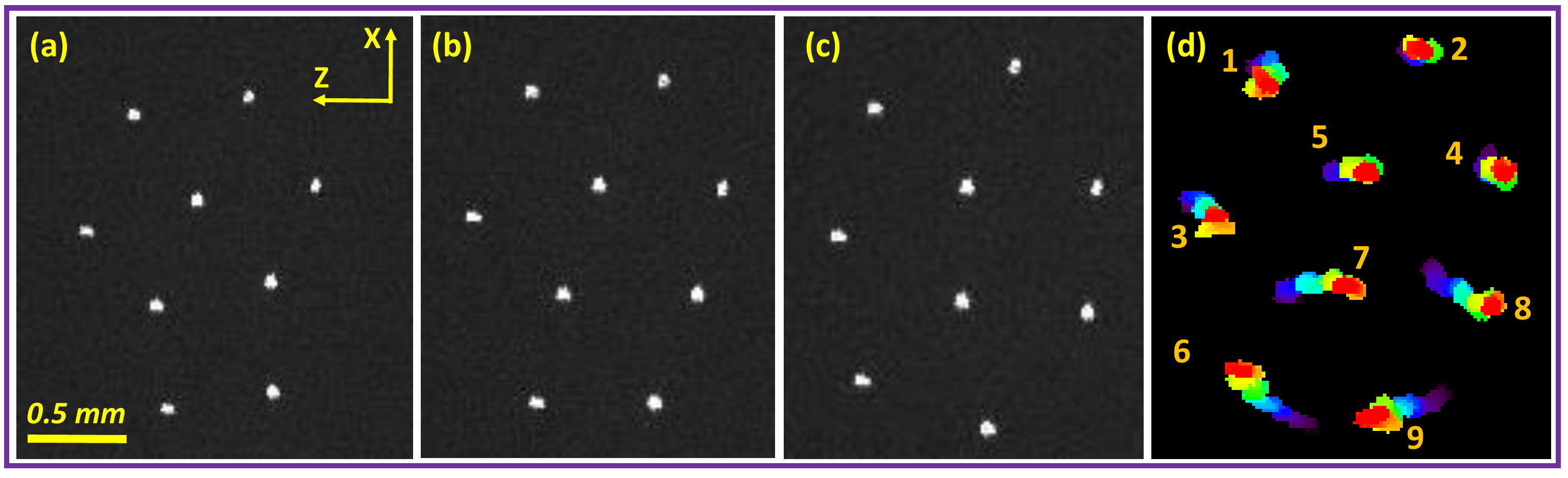}
\caption{\label{fig:fig6} (a) Camera image of cluster with N=9 in meta-stable state,  (b) Cluster during the transition to ground state  (c) Cluster image in the ground state and (d) Overlapped position coordinates during the transition (Colors represent the progress of  time with blue representing the initial time and red the final time.).}
\end{figure}
\begin{figure}[ht]
\includegraphics[scale=1.0]{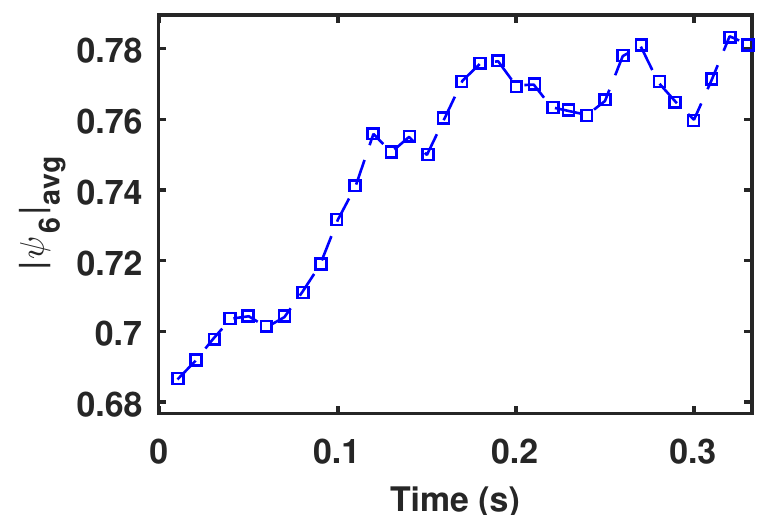}
\caption{\label{fig:fig7}Time evolution of the average local orientational order parameter ($|{\psi_6}|_{avg}$) of two particles (particles 5 and 7 as shown in Fig.~6), during the self-organization. }
\end{figure}
\begin{figure*}[ht]
\includegraphics[scale=1.0]{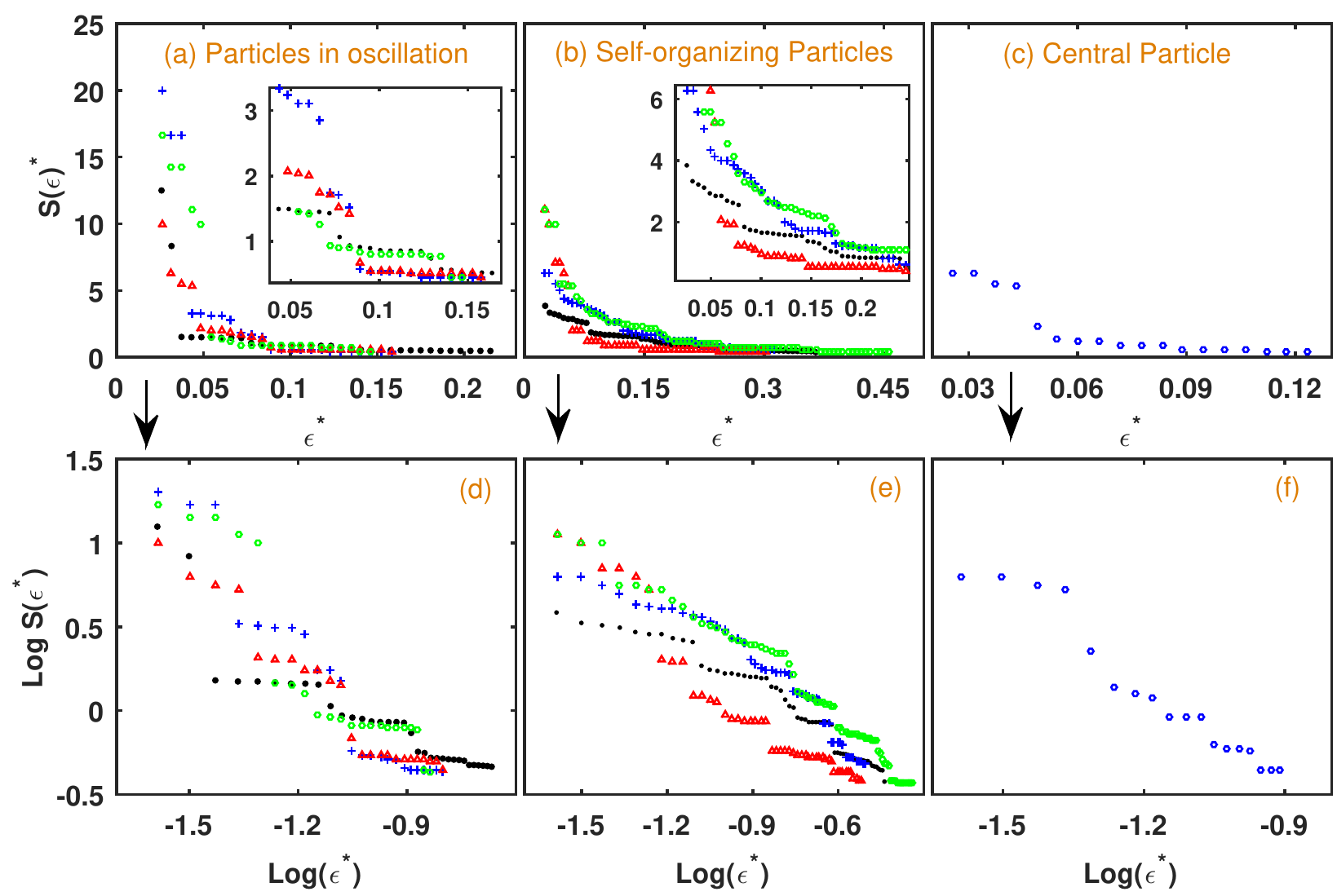}
\caption{\label{fig:fig8} Variation of dynamical entropy with normalized radius for the particles which (a) are in oscillation around their equilibrium position, (b) are in the process of self-organization to form hexagonal symmetry in the ground state and (c) are confined at the centre of the cluster and oscillate around its equilibrium position. The insets in Fig.~8(a) and Fig.~8(b) show zoomed views of the regions where the data points overlap. (d), (e), and (f) present the variation of dynamic entropy on a logarithmic scale with the normalised radius for the particles described in Fig.~8(a), (b) and (c), respectively.} 
\end{figure*}
%
Fig.~\ref{fig:fig6} shows the time evolution of different states of a typical cluster from its meta-stable state to the ground state.  Fig.~\ref{fig:fig6}(a) depicts the initial metastable state which finally transit to the ground state as shown in Fig.~\ref{fig:fig6}(c) though an intermediate state (see Fig.~\ref{fig:fig6}(b)) for N=9. Fig.~\ref{fig:fig6}(d) depicts the overlapping positions of particles during this time evolution in which the blue color represents the initial position whereas the red is for the final position. The dust cluster is observed to be self organizing to achieve the maximum hexagonal symmetry with time.  Fig.~7 depicts the time evolution of the average local orientational order parameter ($|{\psi_6}|_{avg}$) of two particles (particles 5 and 7 as shown in Fig.~6), which reside inside the cluster during the self-organization. $|{\psi_6}|_{avg}$ is found to be around 0.68 at the time when two particles are added to a cluster with seven particles. Over time, the particles self-organize and as a result $|{\psi_6}|_{avg}$ increases to a value of 0.78. The increase in $|{\psi_6}|_{avg}$ is attributed to the self-organization of the particles 6, 7, 8 and 9 as shown in  Fig.~6, which try to acquire a better hexagonal symmetry around the seventh particle. This essentially signifies that the complex plasma cluster system always prefers a hexagonal symmetry for its unit cell and tries to self-organize to such a configuration. In case of an inadequate number of particles to form perfect hexagonal cells, the system acquires maximum hexagonal symmetry from the available number of particles in the cluster. \par 

 Interestingly, the particles situated at the bottom of the dust cluster are particularly active during the transition as these particles are less hexagonally oriented compared to the particles situated at the top of the cluster in the initial stage. Specifically four particles in the bottom are seen to actively  rearrange their  positions to attain a stable ground state structure.  In order to quantify the individual particle dynamics during the cluster de-excitation, the dynamic entropy \cite{dynamicentropyref,dynamicentropy} for each particle is estimated while it self organizes. Dynamic entropy, $S(\epsilon)$ of a particle is estimated as the inverse of the first passage time $\tau(\epsilon)$, where $\tau(\epsilon)$ is the time taken for a particle to cross the boundary of a circle with radius $\epsilon$ in which the particle was at the centre of the circle at t=0. In brief, $S(\epsilon)=1/\tau(\epsilon)$ and $\tau(\epsilon)=\int_{0}^{\infty} P_\epsilon(t) dt$, where $P_\epsilon (t)$ is the probability that the particle reaches the distance $\epsilon$  in the time interval of $t$ and $t+dt$. \par

 In our experimental analysis, dynamic entropy of each dust particle is estimated for the nine particle cluster. For each particle, an imaginary circle of radius $\epsilon^*$ is drawn around it and the first passage time is estimated by analyzing the trajectories during the de-excitation process to determine the dynamic entropy. Fig.~\ref{fig:fig8} shows the dependence of dynamic entropy of individual particles with $\epsilon^*$, where $\epsilon^*=\epsilon/d$ ($d$ being the inter-particle distance). Fig.~\ref{fig:fig8}(a), (b) and (c) represent the dynamic entropy of different sets of particles for a given cluster. The insets in Fig.~\ref{fig:fig8}(a) and Fig.~\ref{fig:fig8}(b) show zoomed views of the regions where the data points overlap. Fig.~\ref{fig:fig8}(d)--Fig.~\ref{fig:fig8}(f) depict the same plots in log scale so that the trajectories of Fig.~\ref{fig:fig8}(a)--Fig.~\ref{fig:fig8}(c) can be distinguished. It is observed that the dynamic entropy and corresponding range of $\epsilon^*$ is different for different particles during the de-excitation. From the analysis it is found that, out of nine particles, four particles are having entropy in same range of $\epsilon^*$, another four particles in another range and a single particle exhibits different $\epsilon^*$ range. Four particles are found to have finite values of entropy upto $\epsilon^*\sim 0.15-0.2$ which is depicted in Fig.~\ref{fig:fig8}(a). This essentially indicates that this set of particles moves a distance of 0.15-0.2 times inter-particle distance during the rearrangement in the cluster. After this distance ($\epsilon^*\sim 0.15-0.2$), the corresponding first passage time will be infinity and hence the dynamic entropy goes to zero. The dynamic entropy of these four particles as shown in Fig.~\ref{fig:fig8}(a) corresponds to the particles 1, 2, 3 and 4 of Fig.~\ref{fig:fig6}(d), which are in lattice oscillation around their equilibrium positions. Furthermore, Fig.~\ref{fig:fig8}(b) shows that another four set of particles whose dynamic entropies are measured  for higher values of $\epsilon^*\sim 0.3-0.45$. These particles possess comparatively higher mobility than that of previous set of particles as shown in Fig.~\ref{fig:fig8}(a). These particles are identified as 6, 7, 8 and 9 in Fig.~\ref{fig:fig6}(d) since they move from their equilibrium position and rearranging to form the ground state. However, the dynamic entropy of one particle is measured only up to $\epsilon^*\sim 0.12$ (see in Fig.~\ref{fig:fig8}(c)). This is the particle which is located at the center of the cluster (particle 5 in Fig.~\ref{fig:fig6}(d)) with a somewhat good hexagonal symmetry around it and is confined by surrounding particles. Hence it is  not able to move much from the mean position. Therefore, by analyzing the particle trajectory as shown in Fig.~\ref{fig:fig6}(d) and the dynamic entropy measurements as in shown in Fig.~\ref{fig:fig8},  the dynamics of individual dust particles are explored during the transition from excited to ground state.

\section{Conclusion}\label{sec:conclusion}
To conclude, classical complex Coulomb clusters are produced in the background of a DC glow discharge Argon plasma in DPEx device in which mono-disperse spherical dust particles are employed as dust grains. The cluster size is varied by adding different number of particles as N=9, 14, 16, 18, 19 and 28. In order to analyze the structural details of the cluster, the hexagonal symmetry is quantified using the local orientational order parameter ($|{\psi_6}|$). The hexagonal symmetry is found to be very sensitive to the number of particles. It is found that the system has high stability and maximum hexagonality when N=18. Interestingly, the system becomes disordered when one particle is further added to a cluster of 18 particles since the number of particles are sufficient to form hexagonal cells by self organizing in the given potential well. Moreover, the thermodynamics of the cluster for different number of particles is examined by estimating the screened Coulomb coupling parameter using Langevin dynamics.  Similar to hexagonality, thermodynamics of the dust cluster is also found to be sensitive to the number of particles. Coupling parameter is found to show a high value at N=18 where the system is dominated with hexagonal symmetry and reduces with the deterioration in the hexagonal symmetry at N=19. This experimental observation suggests an intimate link between the configurational ordering and the thermodynamics of a cluster system - an insight that might be of practical value e.g.  in controlling the dynamics of micro and nano-sized colloidal particle assemblies used in the fabrication of functional materials.  In addition, the self-organization of a cluster from a meta-stable state to the ground state is investigated using the dynamic entropy method for a cluster with N=9. In this process, the system is found to acquire a state with higher hexagonal symmetry in ground state compared to the metastable state. Dynamics of individual components of the dust cluster is also tracked by estimating the dynamic entropy. Four particles are found to be actively re-arranging themselves in the region where hexagonal ordering is low. Other particles are found to be oscillating around their lattice positions with a minor self-organization. The dynamic entropy measurement can be extended to other problems like investigating individual particle dynamics during phase transitions or in phase co-existence states. Our findings may thus prove useful in promoting the practical utility of these thermodynamics based diagnostic tools for investigating strongly coupled finite systems.
\\\\

\noindent \textbf{ACKNOWLEDGMENTS}\\\\
A.S. is thankful to the Indian National Science Academy (INSA) for their support under the INSA Senior Scientist Fellowship scheme. MGH acknowledges Ms. Swarnima Singh for her contribution.\\ \\
\noindent \textbf{DATA AVAILABILITY}\\\\
The data that support the findings of this study are available from the corresponding author upon reasonable request.\\\\
\noindent  \textbf{References}
\bibliography{referencecluster}
\end{document}